\documentclass[11pt]{article}
\usepackage[utf8]{inputenc}
\usepackage{amsmath,amssymb,bbm}

\newcommand{\bs}{\boldsymbol}
\newcommand{\diff}{\mathrm{d}}

\allowdisplaybreaks[4]

\usepackage{algorithm}
\usepackage{algorithmicx}
\usepackage{algpseudocode}

\usepackage[whole]{bxcjkjatype}

\usepackage{graphicx}[dvipdfmx]

\usepackage{mdframed}
\usepackage{natbib}
\usepackage{hyperref}
\hypersetup{
    colorlinks = true,
	citecolor=blue,
	linkcolor=blue
}

\usepackage{geometry}
\usepackage{enumerate}
\usepackage{amsthm}
\theoremstyle{definition}

\newtheorem*{theorem*}{Theorem}

\newtheorem*{note*}{Note}

\usepackage{fourier}
\usepackage{subcaption}

\newcommand{\X}{\mathsf{X}}
\newcommand{\Y}{\mathsf{Y}}
\newcommand{\U}{\mathsf{U}}
\newcommand{\V}{\mathsf{V}}
\newcommand{\setX}{\mathcal{X}}
\newcommand{\setY}{\mathcal{Y}}

\newcommand{\hr}{\widehat{r}}
\newcommand{\hG}{\widehat{G}}
\newcommand{\hH}{\widehat{H}}

\newcommand{\tK}{\widetilde{K}}
\newcommand{\eqdist}{\overset{\mathcal{L}}{=}}

\newcommand{\jac}{J}

\usepackage{authblk}

\title{A multivariate adaptation of \\ 
direct kernel estimator of density ratio}
\author[1,2]{Akifumi Okuno\thanks{okuno@ism.ac.jp; A. Okuno was supported by JSPS KAKENHI (21K17718, 22H05106)}}
\affil[1]{The Institute of Statistical Mathematics}
\affil[2]{RIKEN Center for Advanced Intelligence Project}
\date{\empty}

\begin{document}

\maketitle

\begin{abstract}
\citet{cwik1989estimating} introduced a univariate kernel density ratio estimator, which directly estimates the ratio without estimating the two densities of interest. This study presents its straightforward multivariate adaptation.
\end{abstract}


\section{Introduction}

Let $d \in \mathbb{N}$, and let $\bs \X=(\X_1,\X_2,\ldots,\X_d),\bs \Y=(\Y_1,\Y_2,\ldots,\Y_d) \in \mathbb{R}^d$ be independent random variables. 
These random variables follow two $d$-variate distributions $F$ and $G$, whose probability density functions are $f$ and $g$. 
With access to i.i.d. $n$ copies of $\bs \X \sim F$ denoted as $\setX=\{\bs X_1,\bs X_2,\ldots,\bs X_n\}$ (and $\setY=\{\bs Y_1,\bs Y_2,\ldots,\bs Y_m\}$ for $\bs \Y \sim G$, respectively), this study explores a nonparametric estimator for the density ratio
\[
    r(\bs z)=\frac{f(\bs z)}{g(\bs z)}.
\]

Density ratio plays a crucial role in various statistical problems, particularly in transfer learning. 
See, e.g., \citet{pan2009survey}, \citet{weiss2016survey} and \citet{zhuang2020comprehensive} for a comprehensive survey, where the consideration of the covariate shift~\citep{shimodaira2000improving} is essential. 
For instance, for regression problems, using the loss functions weighted by the density ratio between the training and test covariates, estimated models may exhibit improved performance on the test set, even when the test covariate distribution significantly differs from that of the training data.

One possible approach to estimate the density ratio is to take the ratio between two density estimators. 
Namely, we may consider an estimation consisting of the following two steps: 
(1) we compute two density estimators $\widehat{f},\widehat{g}$ (see, e.g., \citet{tsybakov2009introduction} and \citet{chacon2018multivariate} for nonparametric kernel density estimators), and 
(2) we take their ratio 
\[
\hr_{\text{Indirect}}(\bs z)
:=
\frac{\widehat{f}(\bs z)}{\widehat{g}(\bs z)}.
\]
The above estimator is also referred to as \textit{indirect} density ratio estimator. 
However, depending on the problem setting, the indirect estimator is known to exhibit lower estimation performance compared to \emph{direct} estimators, which directly estimate the ratio $r(\bs z) = f(\bs z) / g(\bs z)$ without separately estimating the two densities $f$ and $g$.

A promising direction to directly estimate the density ratio $r(\bs z)$ is to train a parametric model $r_{\theta}(\bs z)$. 
There exists a variety of previous studies; see, e.g., \citet{qin1998inferences}, \citet{gretton2009covariate}, \citet{sugiyama2007direct}, \citet{yamada2011relative}, \citet{sugiyama2012}, and \citet{kanamori2012statistical} for fundamental approaches, and \citet{rhodes2020telescoping}, 
\citet{kato2021nonnegative}, and \citet{srivastava2023estimating} for further developments including neural network adaptations.

While not currently prevailing, but rather representing a less common approach, this study is motivated to revisit a nonparametric direct kernel density ratio estimator \citep[KDRE;][]{cwik1989estimating}. The asymptotic properties of the direct KDRE were established by \citet{gijbels1995asymptotic} and \citet{chen2009kernel}. 
However, it is worth noting that the estimator is defined exclusively for univariate cases, and it remains unclear whether this estimator can be extended to the multivariate setting. 
Therefore, simply driven by curiosity, this study explores a multivariate adaptation of the direct KDRE.

\subsection{Symbols}
Throughout this paper, we use the unbolded letter $X_{ik}$ to denote the $k$th element of the $d$-variate random variable $\bs X_i=(X_{i1},X_{i2},\ldots,X_{id})$, and $Y_{jk}$ is similarly defined to represent the elements of $\bs Y_j$. 
For univariate case $d=1$, the random variables are simply written by unbolded symbols $X_i,Y_j$. 
While the uppercase letter denotes a random variable, the small letter (e.g., $\bs z \in \mathbb{R}^d$) represents the deterministic variable. 
$\mathbbm{1}(\cdot)$ denotes an indicator function. 
$K:\mathbb{R}^d \to \mathbb{R}$ represents a kernel, which is defined as a non-negative function satisfying $\int K(\bs z) \diff \bs z=1$. Typically, we may employ a boxcar kernel $K(\bs z)=\mathbb{1}(\bs z \in [-1/2,1/2]^d)$.

\section{Direct kernel density ratio estimator (KDRE)}

In this section, we delve into the direct kernel density ratio estimator for the multivariate setting. We begin by explaining the existing univariate estimator in Section~\ref{sec:univariate_KDRE}. Subsequently, we introduce the proposed multivariate adaptation of the estimator in Section~\ref{sec:multivariate_KDRE}.

\subsection{Univariate direct KDRE}
\label{sec:univariate_KDRE}

Instead of the indirect KDRE $\widehat{f}(\bs z)/\widehat{g}(\bs z)$ defined with kernel density estimators $\widehat{f}$ and $\widehat{g}$, 
\citet{cwik1989estimating} introduced a direct KDRE for a univariate setting $d=1$: 
\begin{align}
    \hr_{\text{Direct}}(z)
    :=
    \frac{1}{nh}\sum_{i=1}^{n}
    K\left(
        \frac{\hG(z)-\hG(X_i)}{h}
    \right),
    \quad 
    \hG(z):=\frac{1}{n}\sum_{j=1}^{m}\mathbbm{1}(Y_j \le z). 
    \label{eq:univariate_direct_KDRE}
\end{align}
Direct KDRE estimates the ratio $r(\bs z)=f(\bs z)/g(\bs z)$ without estimating the two densities $f(\bs z),g(\bs z)$ of interest,
and the asymptotic properties of the direct KDRE were studied by \citet{gijbels1995asymptotic} and \citet{chen2009kernel}. 
\citet{motoyama2018direct} proposed a modification that improved its upper bound of the convergence rate. Furthermore, \citet{moriyama2020new} applied a similar technique to hazard ratio estimation. \citet{igarashi2020nonparametric} incorporated the beta kernel to the KDRE to correct the boundary or tail bias.

However, the estimator \eqref{eq:univariate_direct_KDRE} is exclusively defined for univariate cases. To the best of the author's knowledge, there has been no proposed multivariate adaptation of the estimator. In what follows, we first present several flawed attempts at adapting the estimator to the multivariate setting to enhance comprehension, and we present a correct adaptation in Section~\ref{sec:multivariate_KDRE}.

\paragraph{Erroneous multivariate adaptation candidates:} 
The first approach that comes to mind is to replace the cumulative density function $\hat{G}(z)$ in \eqref{eq:univariate_direct_KDRE} with its multivariate counterpart $\hat{G}(\bs z):=\frac{1}{n}\sum_{j=1}^{m}\mathbbm{1}(Y_{j1} \le z_1)\mathbbm{1}(Y_{j2}\le z_2)\cdots\mathbbm{1}(Y_{jd} \le z_d)$. However, this substitution does not effectively extend the estimator, as $\hat{G}:\mathbb{R}^d \to \mathbb{R}$ discards the information in $(d-1)$ dimensions.
Another adaptation approach involves replacing $\hat{G}(z)$ with $\hat{\bs G}(\bs z)=(\hat{G}_1(\bs z),\hat{G}_2(\bs z),\ldots,\hat{G}_d(\bs z))$, using element-wise marginal cumulative density functions $\hat{G}_{\ell}(\bs z):=\frac{1}{n}\sum_{j=1}^{m}\mathbbm{1}(Y_{j\ell} \le z_{\ell})$. However, this substitution also fails to appropriately extend the estimator, as $\hat{\bs G}:\mathbb{R}^d \to \mathbb{R}^d$ neglects correlation information.

\subsection{Multivariate direct KDRE}
\label{sec:multivariate_KDRE}

In this section, we present a multivariate adaptation of the direct KDRE. 
Let $\bs z=(z_1,z_2,\ldots,z_d) \in \mathbb{R}^d$ be a deterministic vector and let $\bs X_i=(X_{i1},X_{i2},\ldots,X_{id})$ and $\bs Y_j=(Y_{j1},Y_{j2},\ldots,Y_{jd})$ be independent copies of the $d$-variate random variables $\bs \X,\bs \Y$, respectively. 
This study presents a multivariate adaptation of the direct KDRE: 
\begin{align}
    \hr_{\text{Direct}}(\bs z)
    :=
    \frac{1}{nh^d} \sum_{i=1}^{n}
    K\left(
        \frac{\bs \hH(\bs z)-\bs \hH(\bs X_i)}{h}
    \right), 
    \quad 
    \bs \hH(\bs z)=(\hH_1(\bs z),\hH_2(\bs z),\ldots,\hH_d(\bs z)),
    \label{eq:direct_KDRE}
\end{align}
where $\hH_{\ell}(\bs z):\mathbb{R}^d \to \mathbb{R}$ is a user-specified estimator for the conditional cumulative density function $H_{\ell}(\bs z)=\mathbb{P}(\Y_{\ell} \le z_{\ell} \mid \Y_1=z_1,\Y_2=z_2,\ldots,\Y_{\ell-1}=z_{\ell-1}):\mathbb{R}^d \to [0,1]$. 
We assume that $\bs \hH(\bs z)$ defined with $\bs \Y$ well estimates $\bs H(\bs z)$ for sufficiently large $m \in \mathbb{N}$. 
For instance, we may employ a kernel estimator for the conditional cumulative function line with \citet{hall1999methods}: 
\begin{align}
    \hH_{\ell}(\bs z)
    =
    \sum_{j=1}^{m}
    w_{\ell,j}(\bs z)
    \mathbbm{1}(Y_{j\ell} \le z_\ell),
    \quad 
    \ell \in \{1,2,\ldots,d\}.
    \label{eq:kcd}
\end{align}
As an instance, $w_{\ell,j}$ can be defined as a Nadaraya-Watson weight:
\begin{align}
    w_{\ell,j}(\bs z)
    =
    \begin{cases}
    1/n & (\ell=1) \\
    \dfrac{\tK_{\ell-1}(\bs \Delta_{j,\ell-1}/\varepsilon_{\ell-1})}{\sum_{j=1}^{m} \tK_{\ell-1}(\bs \Delta_{j,\ell-1}/\varepsilon_{\ell-1})} & (\ell \in \{2,3,\ldots,d\}) \\
    \end{cases},
    \label{eq:NW_weights}
\end{align}
with a user-specified ($\ell-1$)-variate kernel $\tK_{\ell-1}:\mathbb{R}^{\ell-1} \times \mathbb{R}$ and hyperparameter $\varepsilon_{\ell-1}>0$. 
$\bs \Delta_{j,\ell-1} \in \mathbb{R}^{\ell-1}$ denotes the first $\ell-1$ entries of the $d$-variate variable $\bs Y_j-\bs z \in \mathbb{R}^d$, i.e., 
\[
\bs \Delta_{j,\ell-1}=[I_{\ell-1},O_{(\ell-1) \times (d-\ell+1)}](\bs Y_j-\bs z). 
\]
We may employ a Gaussian kernel $\tilde{K}_{\ell-1}(\bs \Delta_{j,\ell-1})=\pi^{-(\ell-1)/2}\exp(-\|\bs \Delta_{j,\ell-1}\|_2^2)$. 
Using \eqref{eq:kcd} and \eqref{eq:NW_weights} (whereby we obtain $\hH_1=\hG$), the proposed direct KDRE \eqref{eq:direct_KDRE} is a natural generalization of the univariate KDRE \eqref{eq:univariate_direct_KDRE}. 
Note that we may employ further extension of the conditional cumulative function estimator; see, e.g., \citet{veraverbeke2013preadjusted} for another nonparametric estimator.

\subsection{Theoretical justification of the multivariate direct KDRE}

This section provides a theoretical justification for the multivariate direct KDRE. 
For brevity, we consider the case that $f,g$ are smooth, and the supports of both functions $f,g$ coincide with $\mathbb{R}^d$. 
For sufficiently large $n,m \in \mathbb{R}^n$ and sufficiently small $\varepsilon_1,\ldots,\varepsilon_{d-1}>0$, the multivariate KDRE \eqref{eq:direct_KDRE} is expected to approximate 
\begin{align}
    \frac{1}{h^d}
    \int K\left(
        \frac{\bs H(\bs z)-\bs H(\bs \X)}{h}
    \right) \diff F(\bs \X)
\label{eq:r_dagger}
\end{align}
by considering the convergence of $\bs \hH(\bs z)$ to $\bs H(\bs z)=(H_1(\bs z),H_2(\bs z),\ldots,H_d(\bs z))$, and the limit of the numerical integration. 
By referring to the conventional kernel density estimation theories~(see, e.g., \citet{tsybakov2009introduction}), \eqref{eq:r_dagger} equipped with a small $h>0$ is expected to approximate the density of the random variable $\bs \V=\bs H(\bs \X)$ at the point $\bs v=(v_1,v_2,\ldots,v_d)=\bs H(\bs z)$, i.e., 
Radon-Nikodym derivative of the probability measure $\mathbb{P}_{\bs \V}$ associated with $\bs \V$ and the Lebesgue measure $\mu$:
\begin{align}
    \frac{\diff \mathbb{P}_{\bs \V}}{\diff \mu}(\bs H(\bs z))
    &\overset{(\star 1)}{=}
    \lim_{\eta \searrow 0}
    \frac{
        \mathbb{P}(\bs V \in B_{\eta}(\bs v))
    }{
        \mu(B_{\eta}(\bs v))
    } \bigg|_{\bs v=\bs H(\bs z)} \nonumber \\
    &\overset{(\star 2)}{=}
    \lim_{\eta \searrow 0}
    \frac{1}{\mu(B_{\eta}(\bs v))}
    \int_{B_{\eta}(\bs v)}
    \frac{f(\bs H^{-1}(\bs u))}{g(\bs H^{-1}(\bs u))}
    \diff \mu (\bs u) 
    \bigg|_{\bs v=\bs H(\bs z)} \nonumber \\
    &\overset{(\star 3)}{=}
    \frac{f(\bs z)}{g(\bs z)}.
    \label{eq:radon_nikodym}
\end{align}
$B_{\eta}(\bs v)=\{\bs v' \in \mathbb{R}^d \mid \|\bs v-\bs v'\| \le \eta\} \subset \mathbb{R}^d$ denotes a ball whose radius is $\eta$ centered at $\bs v$. The equality ($\star 1$) and ($\star 3$) follows from a standard measure theory; see, e.g., Theorem 3 in \citet{evans1992measure} page 38. 
($\star 2$) is obtained by a change of variables; see Appendix~\ref{app:proof} for the proof. 
Therefore, we can expect that the proposed KDRE~\eqref{eq:direct_KDRE} consequently approximates the density ratio $f(z)/g(z)$.



\section{Numerical experiments}

This section provides numerical experiments for $d=2$. Source codes to reproduce the experimental results are provided in \url{https://github.com/oknakfm/MKDRE}.

\subsection{Favorable scenarios}

In the numerical experiments, we employ normal distributions $F=N(\mu_{\bs \X},\Sigma_{\bs \X}),G=N(\mu_{\bs \Y},\Sigma_{\bs \Y})$, where the mean vectors $\mu_{\bs \X},\mu_{\bs \Y}$ and the variance-covariance matrices $\Sigma_{\bs \X},\Sigma_{\bs \Y}$ are specified as
\begin{align*}
    \mu_{\bs \X}
    =
    \begin{pmatrix}
    0 \\ -0.5
    \end{pmatrix},
    \quad
    \Sigma_{\bs \X}
    =
    \begin{pmatrix}
    0.3 & 0.1 \\ 0.1 & 0.3 
    \end{pmatrix},
    \quad 
    \mu_{\bs \Y}
    =
    \begin{pmatrix}
    0 \\ 0
    \end{pmatrix},
    \quad
    \Sigma_{\bs \Y}
    =
    \begin{pmatrix}
    0.5 & 0.1 \\ 0.1 & 0.5 
    \end{pmatrix}.
\end{align*}

We generate samples from the distributions $F$ and $G$ with sizes $n$ and $m$ respectively, and calculate the indirect and direct KDREs using a bandwidth of $h = \varepsilon_1 = 0.1$. We utilize the \verb|kde| function from the \verb|ks| package~\citep{chacon2018multivariate} in the \verb|R| programming language to compute the kernel density estimators.

\begin{figure}[ht]
\centering
\begin{minipage}{0.33\textwidth}
\centering
\includegraphics[width=\textwidth]{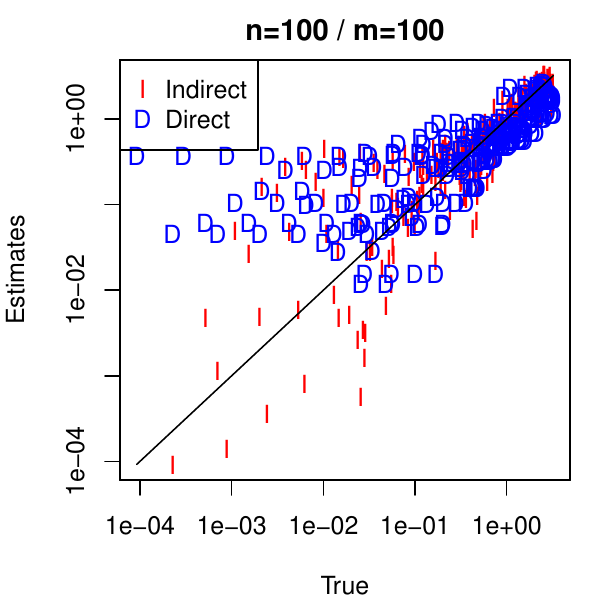}
\subcaption{$n=m=100$}
\end{minipage}
\begin{minipage}{0.33\textwidth}
\centering
\includegraphics[width=\textwidth]{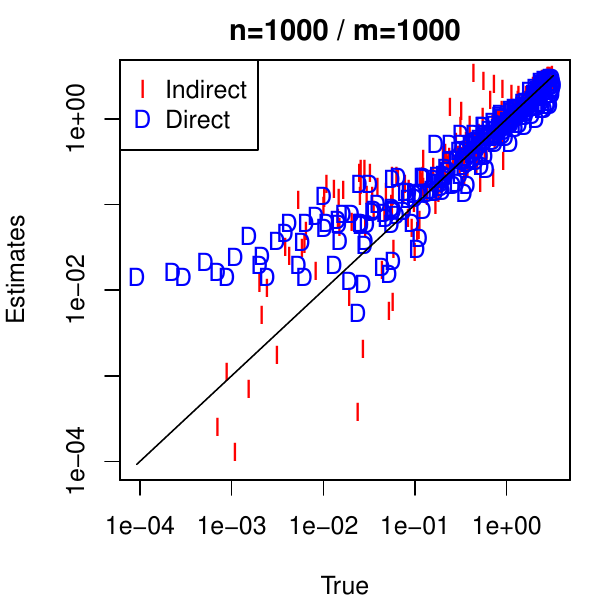}
\subcaption{$n=m=1000$}
\end{minipage}
\begin{minipage}{0.33\textwidth}
\centering
\includegraphics[width=\textwidth]{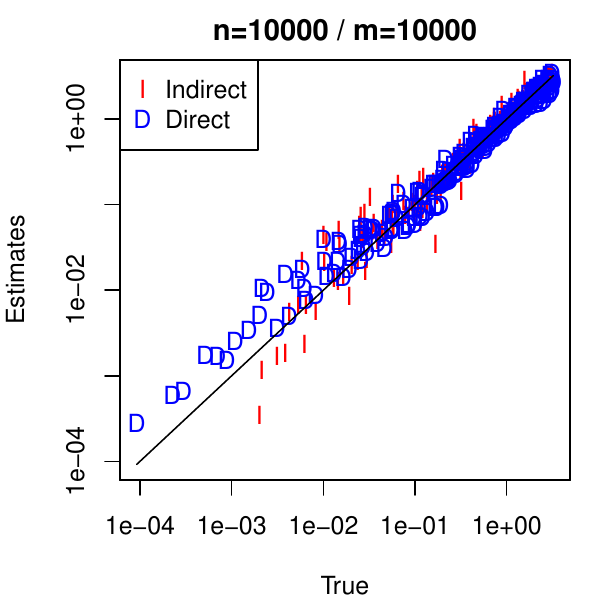}
\subcaption{$n=m=10000$}
\end{minipage}
\caption{True density ratio $r(\bs z)$ and indirect/direct KDREs, for $n=m=100,1000,10000$.}
\label{fig:plot}
\end{figure}

\begin{figure}[ht]
\begin{minipage}{0.33\textwidth}
\centering
\includegraphics[width=\textwidth]{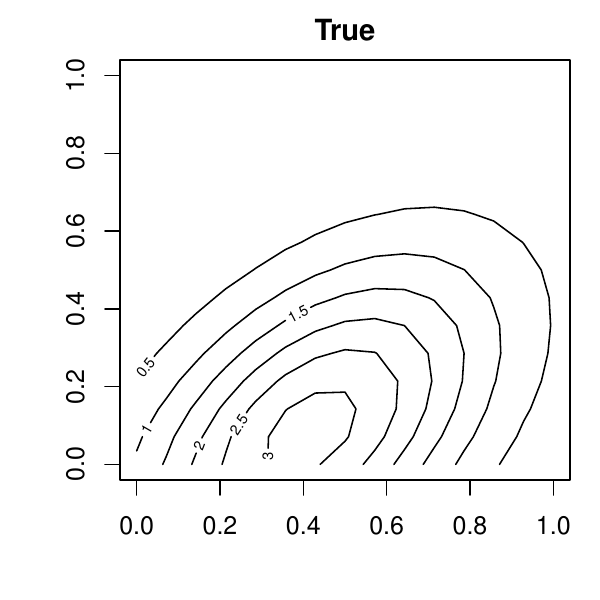}
\subcaption{True}
\end{minipage}
\begin{minipage}{0.33\textwidth}
\centering
\includegraphics[width=\textwidth]{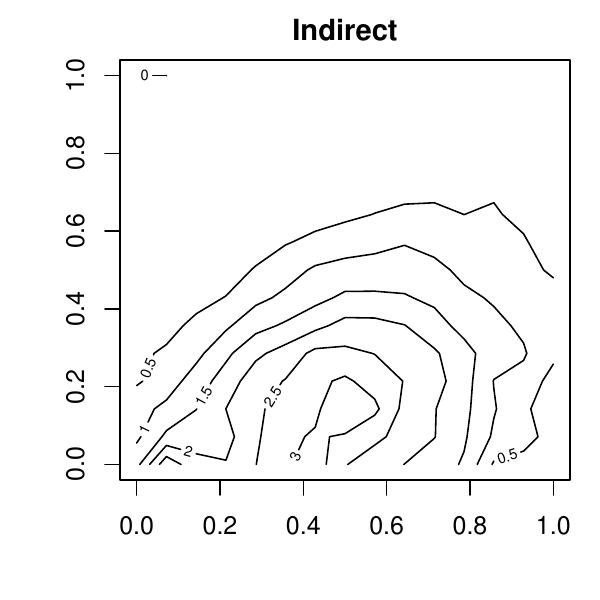}
\subcaption{Indirect}
\end{minipage}
\begin{minipage}{0.33\textwidth}
\centering
\includegraphics[width=\textwidth]{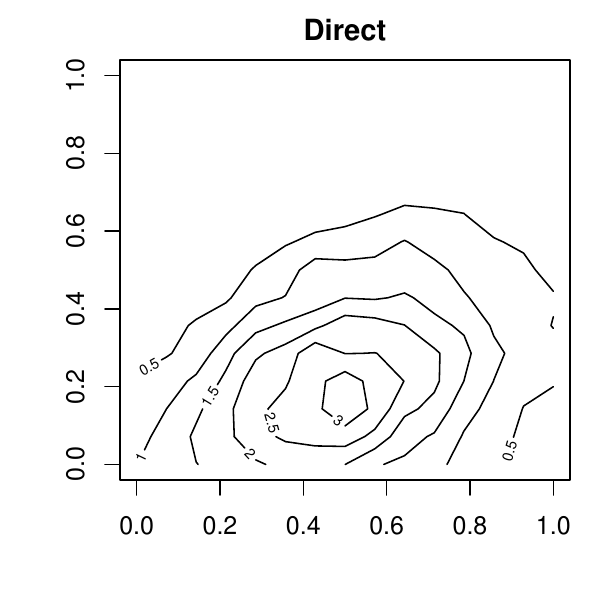}
\subcaption{Direct}
\end{minipage}
\caption{Contour plots of the density ratios for $n=m=10000$.}
\label{fig:contour_plot}
\end{figure}

We create a grid of size $15 \times 15$ over the region $[-1.5, 1.5] \times [-1.5, 1.5]$, and estimate the density ratios at 225 lattice points. The true density ratio, as well as the estimated indirect and direct KDREs computed over these 225 lattice points, are plotted in Figure~\ref{fig:plot} for sample sizes of $n=m=100, 1000, 10000$. 
Specifically, for $n=m=10000$, we also provide contour plots displaying the true density and the indirect/direct KDREs in Figure~\ref{fig:contour_plot}. It is noticeable that the direct KDRE effectively approximates the true density and is comparable to the indirect KDRE.

\subsection{Challenging scenarios}

\begin{figure}[ht]
\centering
\begin{minipage}{0.33\textwidth}
\centering
\includegraphics[width=\textwidth]{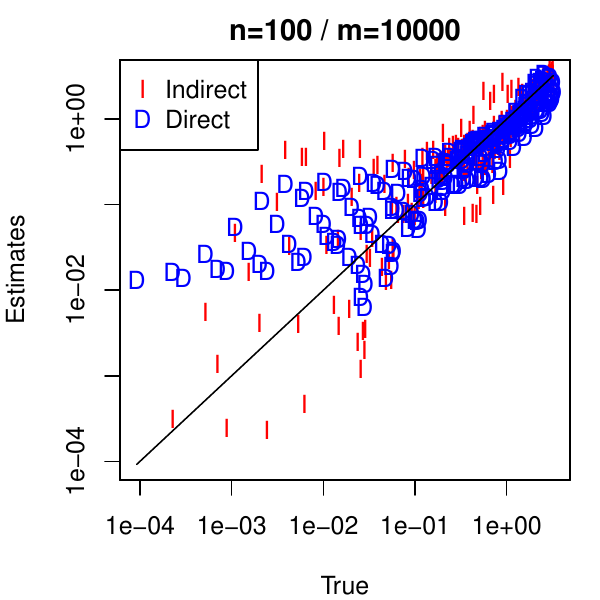}
\subcaption{$n=100,m=10000$}
\label{fig:100_10000}
\end{minipage}
\begin{minipage}{0.33\textwidth}
\centering
\includegraphics[width=\textwidth]{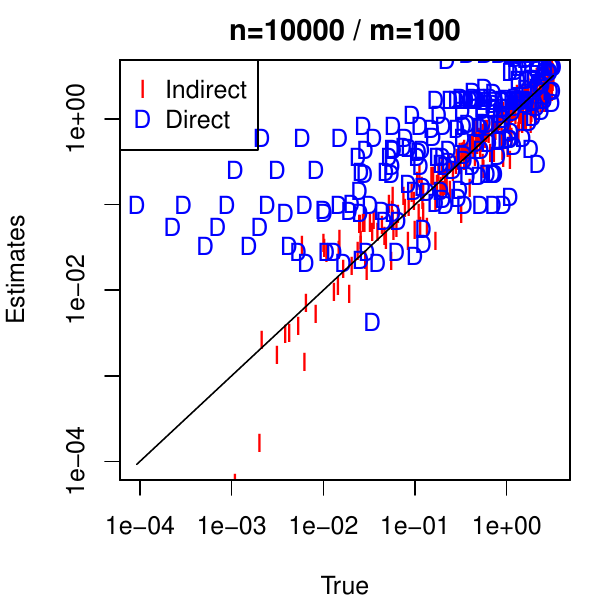}
\subcaption{$n=10000,m=100$}
\label{fig:10000_100}
\end{minipage}
\begin{minipage}{0.33\textwidth}
\centering
\includegraphics[width=\textwidth]{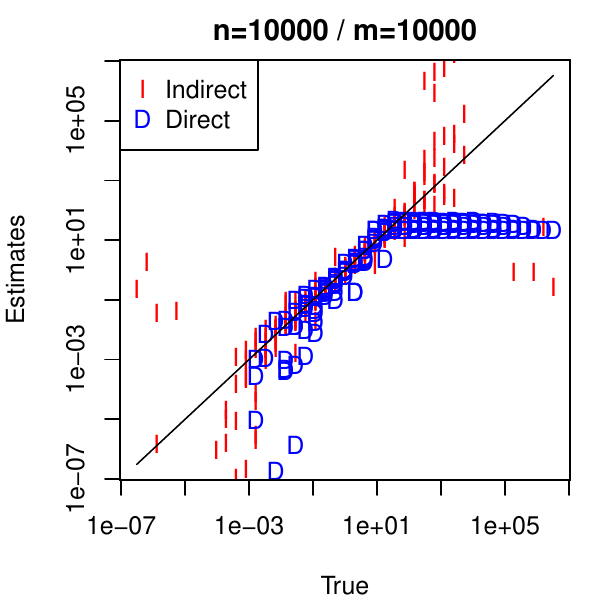}
\subcaption{larger ratio, $n=m=10000$}
\label{fig:larger}
\end{minipage}
\caption{True density ratio $r(\bs z)$ and indirect/direct KDREs.}
\label{fig:plot_additional}
\end{figure}

We also calculate the estimators in challenging scenarios. Figure~\ref{fig:plot_additional}(\subref{fig:100_10000}) and \ref{fig:plot_additional}(\subref{fig:10000_100}) display the estimators in an imbalanced setup: (\subref{fig:100_10000}) with $n=100$ and $m=100000$, and (\subref{fig:10000_100}) with $n=10000$ and $m=100$. 
It is noticeable that the direct estimator exhibits poorer performance when $m$ is relatively small. 

Figure~\ref{fig:plot_additional}(\subref{fig:larger}) illustrates the experimental results with a higher true density ratio. In this experiment, we use normal distributions $F=N(\mu_{\bs \X},\Sigma_{\bs \X})$ and $G=N(\mu_{\bs \Y},\Sigma_{\bs \Y})$ with
\begin{align*}
    \mu_{\bs \X}
    =
    \begin{pmatrix}
    0 \\ -0.5
    \end{pmatrix},
    \quad
    \Sigma_{\bs \X}
    =
    \begin{pmatrix}
    0.2 & 0.1 \\ 0.1 & 0.2 
    \end{pmatrix},
    \quad 
    \mu_{\bs \Y}
    =
    \begin{pmatrix}
    0 \\ 0.5
    \end{pmatrix},
    \quad
    \Sigma_{\bs \Y}
    =
    \begin{pmatrix}
    0.2 & 0.1 \\ 0.1 & 0.2 
    \end{pmatrix}.
\end{align*}
It is evident that the direct estimator struggles to estimate large density ratios. This limitation arises because the direct estimator~\eqref{eq:direct_KDRE} is bounded above by $h^{-d}\sup_{\bs z \in \mathbb{R}^d}|K(\bs z)|$.

\section{Concluding discussion} 
Simply driven by curiosity, this study introduced a multivariate adaptation of the univariate direct KDRE proposed by \citet{cwik1989estimating}. Our experiments demonstrated that the multivariate adaptation yielded comparable results to the indirect KDRE.

For the multivariate case ($d \in \mathbb{N}$), we find no evidence to suggest that the direct KDRE outperforms the indirect KDRE, considering both theoretical and practical perspectives. Unfortunately, the direct KDRE, as considered in this study, suffers more severely from the curse of dimensionality problem (as seen in Figure~\ref{fig:plot_additional}(\subref{fig:10000_100})). This is attributed to its reliance on estimates of the conditional cumulative function $H_{\ell}(\bs z)$, which are generally challenging to estimate stably. Consequently, we do not consider the direct KDRE in its current form to be practical, and we expect further improvements of the multivariate direct KDRE.

\appendix

\section{Change of variables}
\label{app:proof}

This section provides a proof of the equation ($\star 2$) in \eqref{eq:radon_nikodym}. 
In this proof, $\eqdist$ denotes the equality of two random variables in the sense of distribution. 
Consider a function $\bs L(\bs z)=(L_1(\bs z),L_2(\bs z),\ldots,L_d(\bs z))$ defined with the conditional distribution function $L_{\ell}(\bs z)=\mathbb{P}(\X_{\ell} \le z_{\ell} \mid \X_1=z_1,\X_2=z_2,\ldots,\X_{\ell-1}=z_{\ell-1}):\mathbb{R}^d \to [0,1]$. 
$\bs H,\bs L:\mathbb{R}^d \to [0,1]^d$ are invertible functions, and $\bs \X \eqdist \bs L^{-1}(\bs \U)$ holds for the uniform random variable $\bs \U$ defined over the set $[0,1]^d$, by considering the inverse transform sampling. 
The identity $\bs \V \eqdist (\bs H \circ \bs L^{-1})(\bs \U)$ indicates for the jacobian $\jac$ that 
\begin{align*}
\mathbb{P}(\bs \V \in B_{\eta}(\bs v))
&=
\mathbb{P}(\bs \U \in (\bs L \circ \bs H^{-1})(B_{\eta}(\bs z))) \\
&=
\int_{(\bs L \circ \bs H^{-1})(B_{\eta}(\bs z))} \diff \mu(\bs u) \\
&=
\int_{B_{\eta}(\bs v)} |\jac_{\bs L \circ \bs H^{-1}} (\tilde{\bs u})| \diff \mu (\tilde{\bs u}) 
\qquad 
\left(\because \, \tilde{\bs u}:=(\bs H \circ \bs L^{-1})(\bs u) \right)
\\
&=
\int_{B_{\eta}(\bs v)} |\jac_{\bs L}(\bs H^{-1}(\tilde{\bs u}))| \cdot |\jac_{\bs H^{-1}}(\tilde{\bs u})| \diff \mu(\tilde{\bs u}) \\
&=
\int_{B_{\eta}(\bs v)} \frac{|\jac_{\bs L}(\bs H^{-1}(\tilde{\bs u}))|}{|\jac_{\bs H}(\bs H^{-1}(\tilde{\bs u}))|} \diff \mu(\tilde{\bs u}) \\
&=
\int_{B_{\eta}(\bs v)} \frac{
    f_1(z_1)f_2(z_2 \mid z_1)f_3(z_3 \mid z_1,z_2) \cdots f_d(z_d \mid z_1,z_2,\ldots,z_{d-1})
    \mid_{\bs z=\bs H^{-1}(\tilde{\bs u})}
}{
    g_1(z_1)g_2(z_2 \mid z_1)g_3(z_3 \mid z_1,z_2) \cdots g_d(z_d \mid z_1,z_2,\ldots,z_{d-1})
    \mid_{\bs z=\bs H^{-1}(\tilde{\bs u})}
} \diff \mu (\tilde{\bs u}) \\
&=
\int_{B_{\eta}(\bs v)}
    \frac{f(\bs H^{-1}(\tilde{\bs u}))}{g(\bs H^{-1}(\tilde{\bs u}))}
    \diff \mu (\tilde{\bs u})
\end{align*}
$f_{\ell}(z_{\ell} \mid z_1,z_2,\ldots,z_{\ell-1})$ denotes the conditional probability density function associated with $H_{\ell}$, and $g_{\ell}$ is similarly defined for $L_{\ell}$. The assertion is proved. 
\qed

\bibliographystyle{apalike}
\bibliography{kdre}

\end{document}